\documentclass[]{interact}

\usepackage{epstopdf}
\usepackage[caption=false]{subfig}
\usepackage{graphicx}

\usepackage[numbers,sort&compress]{natbib}
\bibpunct[, ]{[}{]}{,}{n}{,}{,}
\renewcommand\bibfont{\fontsize{10}{12}\selectfont}
\makeatletter
\def\NAT@def@citea{\def\@citea{\NAT@separator}}
\makeatother

\theoremstyle{plain}

\theoremstyle{definition}

\theoremstyle{remark}

\begin{document}

\articletype{}

\title{Venus, Phosphine and the Possibility of Life}

\author{
\name{David L Clements\textsuperscript{a}\thanks{CONTACT David L. Clements Email: d.clements@imperial.ac.uk}}
\affil{\textsuperscript{a}Department of Physics, Imperial College, Prince Consort Road, London, SW7 2AZ, UK}
}

\maketitle

\begin{abstract}
The search for life elsewhere in the universe is one of the central aims of science in the 21st century. While most of this work is aimed at planets orbiting other stars, the search for life in our own Solar System is an important part of this endeavour. Venus is often thought to have too harsh an environment for life, but it may have been a more hospitable place in the distant past. If life evolved there in the past then the cloud decks of Venus are the only remaining niche where life as we know it might survive today. The discovery of the molecule phosphine, PH$_3$, in these clouds has reinvigorated research looking into the possibility of life in the clouds. In this review we examine the background to studies of the possibility of life on Venus, discuss the discovery of phosphine, review conflicting and confirming observations and analyses, and then look forward to future observations and space missions that will hopefully provide definitive answers as to the origin of phosphine on Venus and to the question of whether life might exist there.

\end{abstract}

\begin{keywords}
Venus; astrobiology; search for life; 
\end{keywords}

\section{Introduction}

The search for life elsewhere in the universe is one of the major driving forces for astronomy and astrophysics in the 21st century \cite{astro2020} with extremely large telescopes on the ground and in space being designed and built to this end. The main goal of these facilities is to study the atmospheres of rocky, terrestrial planets in orbit around other stars - so-called exo-planets - and follows the explosive development of exoplanet studies since their first discovery in 1995 \cite{mq95}. We now know of over 5000 exoplanets, and more are announced all the time (for the latest information on exoplanet discoveries see \verb|www.exoplanet.eu|). However, while convincing evidence for life on exoplanets may arrive in the next 20 years, there are many questions that such observations will not be able to answer, including the biochemical processes exoplanet life uses, how and when it originated, and how it evolved into whatever form it has today - a form that will also be unknowable.

If we want answers to these further questions the only places they may be answered is in our own Solar System. While still very distant, planets such as Mars, the Jovian moon Europa, and Saturn's moons Enceladus and Titan, which might harbour signs of life, are accessible to direct {\em in situ} studies that can answer these questions. And any answers inevitably lead back to the question of our own origins on Earth, as well as the more general issue of the prevalence of life in the universe.

The search for signs of life on Mars is well underway, with orbiting spacecraft looking at its atmosphere (eg. the ExoMars Trace Gas Orbiter (TGO) \cite{vago15}) and an ever-increasing number of rovers scouring its surface \cite{f20}. Some of these are already preparing samples of rock to be returned to Earth for laboratory analysis. Further out in the Solar System, the first stages of the exploration of Jupiter's moons Ganymede and Europa, in part for signs of habitable environments, are already under development with the European Space Agency's (ESA's) JUpiter ICy moons Explorer (JUICE) \cite{g13}, aimed primarily at Ganymede, due for launch in 2023, and NASA's Europa Express, aimed squarely at Europa, due for launch in 2024. Plans are at an earlier stage for the exploration of the moons of Saturn that might harbour life, including Titan with its thick atmosphere and Enceladus with its plumes of water vapour spewing into space, but it is clear that these too will be visited sometime in the next few decades.

Until very recently this list of targets - Mars, Europa, Ganymede, Titan and Enceladus - would have been considered the most likely places to find signs of life in the Solar System. It was thus rather surprising to find Venus added to this list in late 2020 with the discovery of an unusual gas, phosphine, chemical formula PH$_3$, in its atmosphere \cite{g21}. Venus, as we will see below, has a surface that is completely hostile to life and so had been largely discounted from these considerations. But, as we will also see, there have long been thoughts that there might be niches in the atmosphere of this planet that might be more favourable to the existence of life than its deeply unpleasant surface \cite{ms67, l18}. In this article we discuss how life is sought using atmospheric observations, why phosphine is a potential signature of biological activity, how it was detected on Venus, and what its discovery might mean for our understanding of the history of Venus and of life in the Solar System. We also look at the prospects for future studies of Venus in search of further possible signs of life.

\section{What is Life?}

The formal definition of life adopted by NASA for searches for life elsewhere is that `Life is a self-sustaining chemical system capable of Darwinian evolution' \cite{b10}. This definition clearly applies to most things that we would consider alive on Earth, but it does leave some things out. Viruses, for example, cannot reproduce on their own, but instead require a host cell, whose reproductive machinery they take over. Alternative definitions are available (eg. \cite{v19}), but the NASA definition seems a useful starting point to begin any discussion of where life might exist in the Solar System or elsewhere. Given this definition we can start to examine what the requirements might be for life to exist.

From a physicist's perspective the key thing that life needs to be able to support itself is a source of energy. For most of the life we are familiar with on the Earth that source of energy is the Sun - plants photosynthesise using sunlight, while animals and other organisms consume plants in various ways, including eating things that have eaten plants. However, sunlight is not the only source of energy used by life on Earth. In the depths of Earth's oceans, where sunlight never shines, there are thriving communities of organisms surrounding and fed by hydrothermal vents \cite{l93}. These arise where ocean water can  enter the Earth's crust and be heated by magma. The heated water dissolves minerals from the rocks and circulates back into the ocean through the vents. The primary producers of energy in these vents are chemosynthetic bacteria that use a variety of processes to derive energy from the chemicals emerging from the vents. These then support a diverse community of other organisms around the vents, including giant tube worms that can be up to 3 metres long. Hydrothermal vents in the young Earth are a possible site for the first emergence of life on our planet \cite{d17}. Similar hydrothermal vent structures are thought to exist beneath the ice-covered surfaces of moons like Europa and Enceladus \cite{h20}. However, life without light on Earth is not limited to hydrothermal vents. It has even been suggested that most ecosystems on Earth exist in the dark, deriving their energy from chemical processes separate from, and independent of, photosynthesis \cite{ed12}. This clearly has implications for the search for life elsewhere.

What requirements for life are there beyond a source of energy? Revisiting NASA's definition of life we see that it is defined as a {\em chemical} system. The chemical basis for all the life we know on Earth is the element carbon.  This element is exceptional in the Periodic Table as the lightest element in Group IV.  It thus has a half-filled electron shell giving it a valence of 4 so it can donate or receive up to 4 electrons, allowing it to bind with itself to form chains, and with numerous other elements. Some of the commonest are hydrogen, oxygen and nitrogen, contributing to the wide variety of complex organic compounds found in living organisms. The next heaviest Group IV element is silicon. It has similar high valence and has been proposed as an alternative building block for life \cite{p20}, but there are a number of issues which mean that carbon is a better choice.

Given a chemical basis for life, there must be a way for the necessary chemical reactions to take place so that the processes of life can operate. The best way to achieve this is for the reacting chemicals to be dissolved by some solvent so that they can easily combine and interact. On Earth the solvent behind all biology is water. While other solvents have been suggested, especially in environments too hot or too cold for liquid water to be available \cite{b04}, water remains the only solvent which we know is associated with life. The bulk of our searches for life elsewhere have thus been focussed on places where liquid water might, or is known to, exist.

Our consideration of the nature of life thus leads us to the conclusion that we should be looking for life on  an astronomical body where there is a ready supply of energy, where complex carbon-based chemistry can take place, and where liquid water can exist. In our own Solar System this leads us to focus on the planet Mars and the icy moons of gas giants, such as Europa and Ganymede orbiting Jupiter, and Titan and Enceladus orbiting Saturn. Venus does not appear on this list because, as we will see below, its current surface conditions are far too hostile for life as we know it, but, as we will also see, this may not always have been the case, and there may be loopholes that could allow any life that might have emerged on the surface of this planet to persist today in the dense clouds above its surface.

\section{Looking for Life}

Life is easy to find on Earth - we are surrounded by it - but looking for life on astronomical bodies, even our closest neighbours, is a rather more difficult task, especially since we do not expect to find something as obviously alive as a tree or a cow. In the absence of complex macroscopic life we have to look for signs of activity from microbial life. What might these be? Using Earth as an example, the clearest sign that life exists is the presence of oxygen in our atmosphere. This is generated by photosynthesising organisms. Today we would associate this with plants, but in the ancient history of our planet, the first oxygenating organisms were single-celled microbes known as blue-green algae. These were responsible for the Great Oxygenation Event which took place about 2.5 billion years ago \cite{l14}. Before this event oxygen was not a major constituent of Earth's atmosphere. After it, the Earth had an atmosphere much closer to what we see today, with abundant free oxygen available across the planet. In principle, the presence of oxygen in the atmosphere of Earthlike exoplanets will be detectable by future instruments through the absorption of ozone, O$_3$, at mid-infrared wavelengths \cite{f21}. 

Molecules that potentially indicate the presence of life, such as oxygen and ozone, are known as biosignatures (see \cite{j21} and references therein). They include oxygen, ozone, methane (CH$_4$), N$_2$O, C$_2$H$_6$ CH$_3$Cl, CH$_3$SH and more. A wider variety of biosignatures than just oxygen and ozone are needed to cope with potentially different biospheres than the one we currently inhabit. In fact, for much of the history of life on Earth, there would not have been sufficient oxygen or ozone for life to be detectable since the Earth was dominated by anaerobic life (ie. life that does not require or produce oxygen). Searches for other small molecules that might be additional biosignatures are also underway \cite{s16}. The common factor among all of these biosignature molecules is that they should not exist in the abundances seen unless biological processes are maintaining their abundance ie. their abundance is out of equilibrium with their environment. However, biological processes are not the only way of maintaining an out-of-equlibrium abundance of some of these biosignatures. For example, the splitting of water into hydrogen and oxygen by stellar ultraviolet radiation, and the subsequent escape of hydrogen from the planet's atmosphere, can produce a significant partial pressure of oxygen on an abiotic (ie. lifeless) planet given certain other conditions \cite{m18}. Care must therefore be taken in interpreting any unusual abundance of a single molecule as an unambiguous biosignature without a good understanding of the broader environment in which it is found.

Phosphine, PH$_3$, is one of the small molecules suggested as a potential biosignature by the team investigating novel biosignature molecules \cite{ss20}. On Earth it is exclusively associated with anaerobic ecosystems, or with human industrial chemistry. Sources have been found associated with anaerobic environments in ponds, marshes and sludges, and specifically with piles of penguin guano and in the faeces of European badgers. While it is clearly associated with anaerobic biology, the specific biochemical pathway for phosphine production in anaerobic systems remains unclear. Its use as a biosignature was originally envisioned in the context of a significant concentration of the gas within the atmosphere of a terrestrial exoplanet with a large anaerobic biosphere. While phosphine has been detected in the vast reducing (ie. hydrogen rich) atmosphere of the gas giant Jupiter \cite{b75}, where it is produced by normal chemical processes in the very dense, hot inner regions then brought to the surface by convection, its presence in the oxidised atmosphere of a terrestrial planet would be difficult to explain by equilibrium chemistry, making it a good candidate biosignature. Observational facilities able to find phosphine in the atmosphere of a distant exoplanet are some way in the future, but, as we see below, a surprise was waiting for us much closer to home.

\section{Venus - an unlikely candidate for astrobiology}

\begin{figure}
\begin{tabular}{cc}
\includegraphics[width=6.85cm]{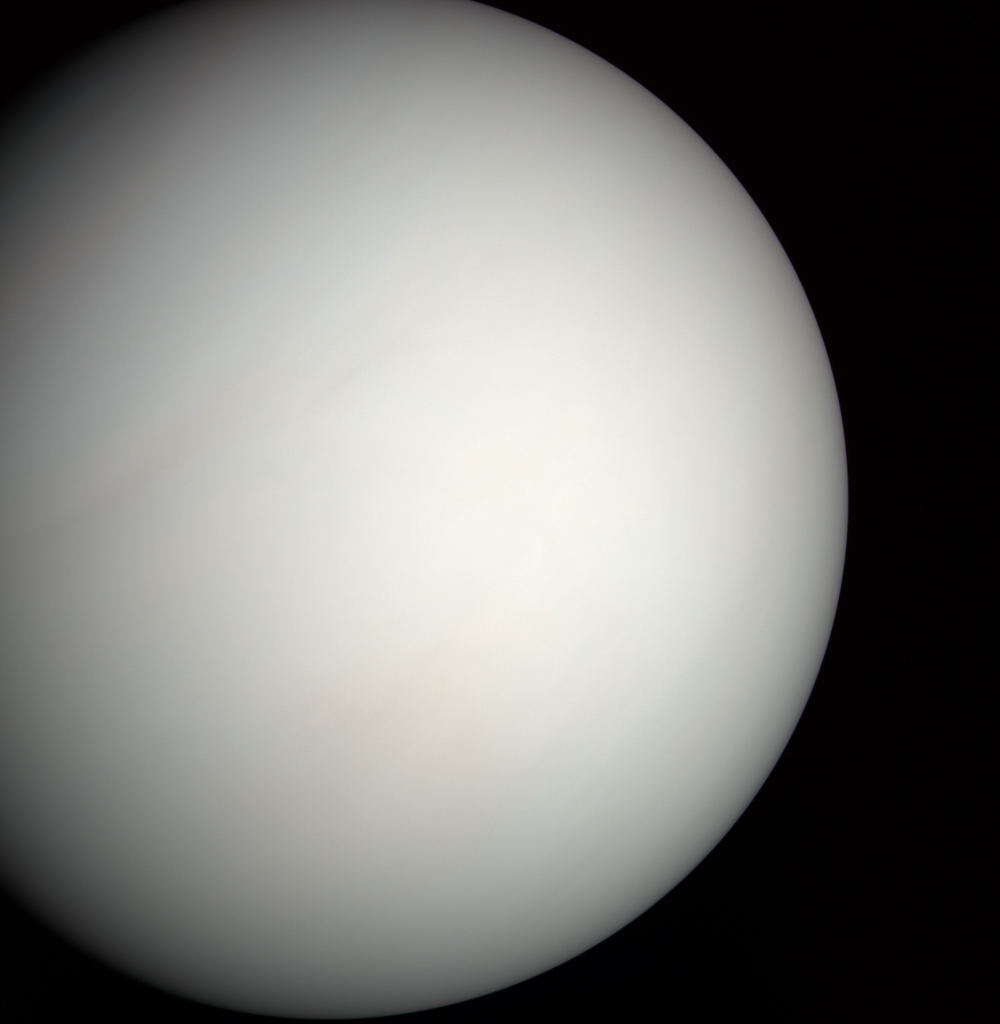}
\includegraphics[width=7cm]{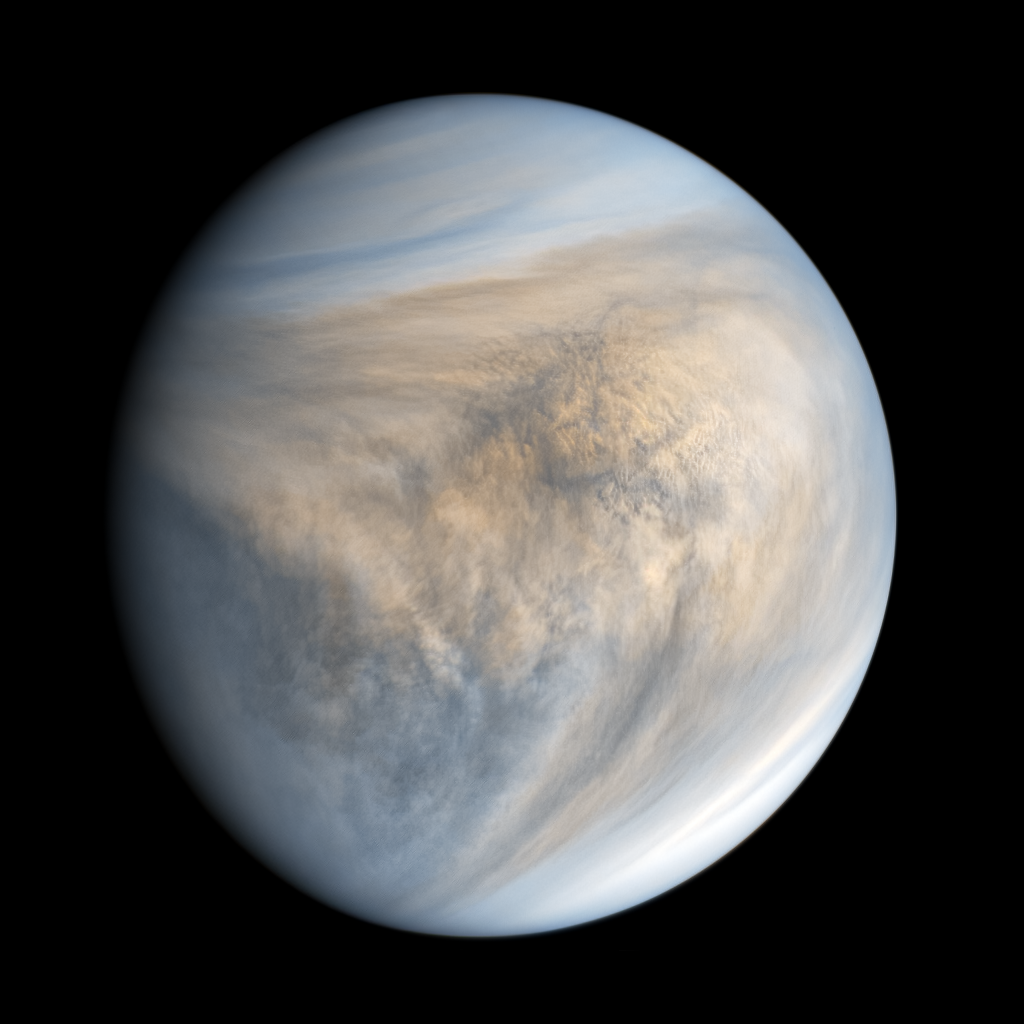}
\end{tabular}
\caption{Left: Venus as seen in the optical by the Messenger mission. In the optical the planet appears almost featureless because of the highly reflective clouds that cover the entire planet. Credit: NASA/Johns Hopkins University Applied Physics Laboratory/Carnegie Institution of Washington Right: Venus as seen in the ultraviolet by the Japanese space mission AKATSUKI. This image is produced by observations in two ultraviolet bands, at 365 and 283 nm. The colours are the result of unexplained ultraviolet absorption by small particles in the cloud layer. Credit: JAXA/ISAS/DARTS/Kevin M. Gill}
\label{fig:venus}
\end{figure}

\subsection{Venus Today}

Venus has often been described as Earth's evil twin since the two planets have very similar sizes and masses, with Venus having a radius about 95\%, and a mass about 82\% of the Earth's. It is also a little under 30\% closer to the Sun than the Earth. 
Viewed from a telescope, Venus appears as a featureless pale disk because it has a thick atmosphere containing a permanent cloud layer, opaque to optical observations, that reflects over 70\% of the sunlight that falls on it (see Figure \ref{fig:venus}). Venus' orbit closer to the Sun, combined with the effects of these reflective clouds, might naively suggest a surface temperature not too different from the Earth. Because of this, in the early 20th century it was thought that the clouds might be water vapour and that the surface of Venus could be habitable, inspiring visions of steamy tropical jungles filled with alien life. Once more detailed observations became available, and space probes started to visit in the 1960s, it became apparent that Venus was very different from these initial speculations.

Rather than having an Earthlike atmosphere, early 1960s space probes like NASA's Mariner 2 and the Soviet Union's Venera 4 found that Venus has an extremely dense atmosphere dominated by carbon dioxide, CO$_2$, with a small amount of nitrogen (3.5\%) and traces of other gases such as sulphur dioxide (SO$_2$). The surface atmospheric pressure on Venus is about 93 times higher than sea level pressure on the Earth. This huge blanket of CO$_2$ absorbs thermal radiation from the surface of the planet which would otherwise be radiated away into space. The Sun's radiation is thus trapped beneath the clouds, leading to a runaway greenhouse effect and surface temperatures of about 735 K (462 C), making it hotter than the maximum surface temperature of Mercury \cite{l04}\footnote{It is worth noting that the discovery of the runaway greenhouse effect on Venus prompted early discussions about the dangers of CO$_2$ emissions from fossil fuel use on Earth and their possible impact on climate.}. To make things even more unpleasant, the thick clouds are largely made up of sulphuric acid as a result of atmospheric SO$_2$ dissolving into droplets of water that condense at altitudes of about 55 km above the surface. All of this combines to make the surface of Venus today an utterly inhospitable place for anything we might consider a biological system. The true nature of the surface of Venus in fact led leading astronomer Carl Sagan to describe the planet as hell.

\begin{figure}
\includegraphics[width=14cm]{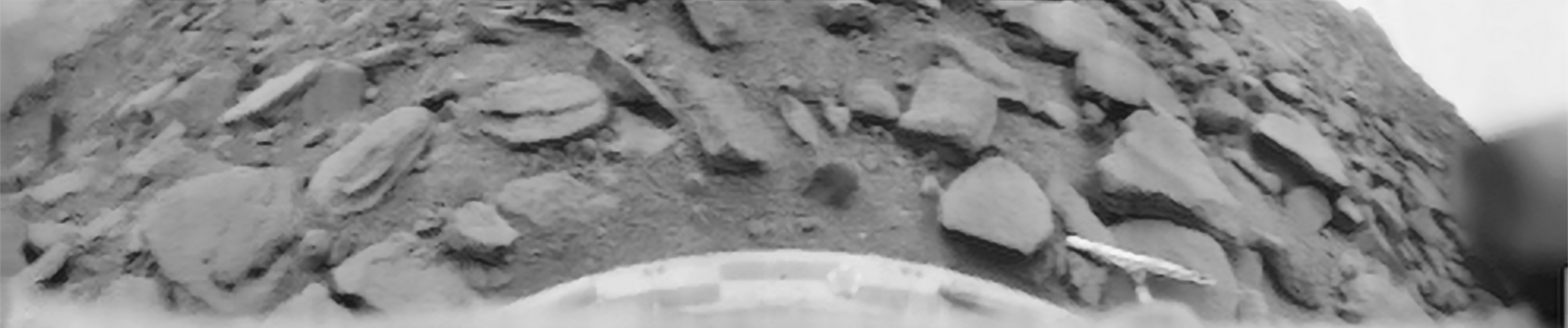}
\caption{A 360 degree panorama of the surface of Venus captured by the Venera 9 probe during its 53 minutes of operation on the harsh surface of the planet. This is the first image ever obtained from the surface of Venus. Part of the probe can be seen at the bottom of the image. }
\label{fig:surface}
\end{figure}

Spacecraft continued to visit Venus after the early missions, including a long series of Venera probes launched by the Soviet Union. Several of these, as well as the Pioneer Venus mission from NASA, sent probes into the atmosphere of Venus to better understand its chemistry and constituents. The Venera and Vega projects also attempted to land probes on the hostile surface of Venus. After a number of failures they were eventually successful and conducted a number of studies. None of the landers lasted very long, with the longest lived surviving only about 2 hours. Nevertheless, some of these landers managed to send back images of the Venusian surface, one of which can be seen in Figure \ref{fig:surface}.

While the surface of Venus is undoubtedly hostile to life at the present time, there is a potential niche for biological processes in the clouds that obscure the planet since they are cool enough for liquid water to be present, and at an atmospheric pressure that matches that of the Earth at sea level \cite{t18, v13} (see Figure 3). Speculation about the possibility of life in the clouds of Venus dates back to the very earliest days of our understanding of the true conditions on the planet \cite{m67} and continues to the present day (eg. \cite{s21, ss21}).

\begin{figure}
\includegraphics[width=14cm]{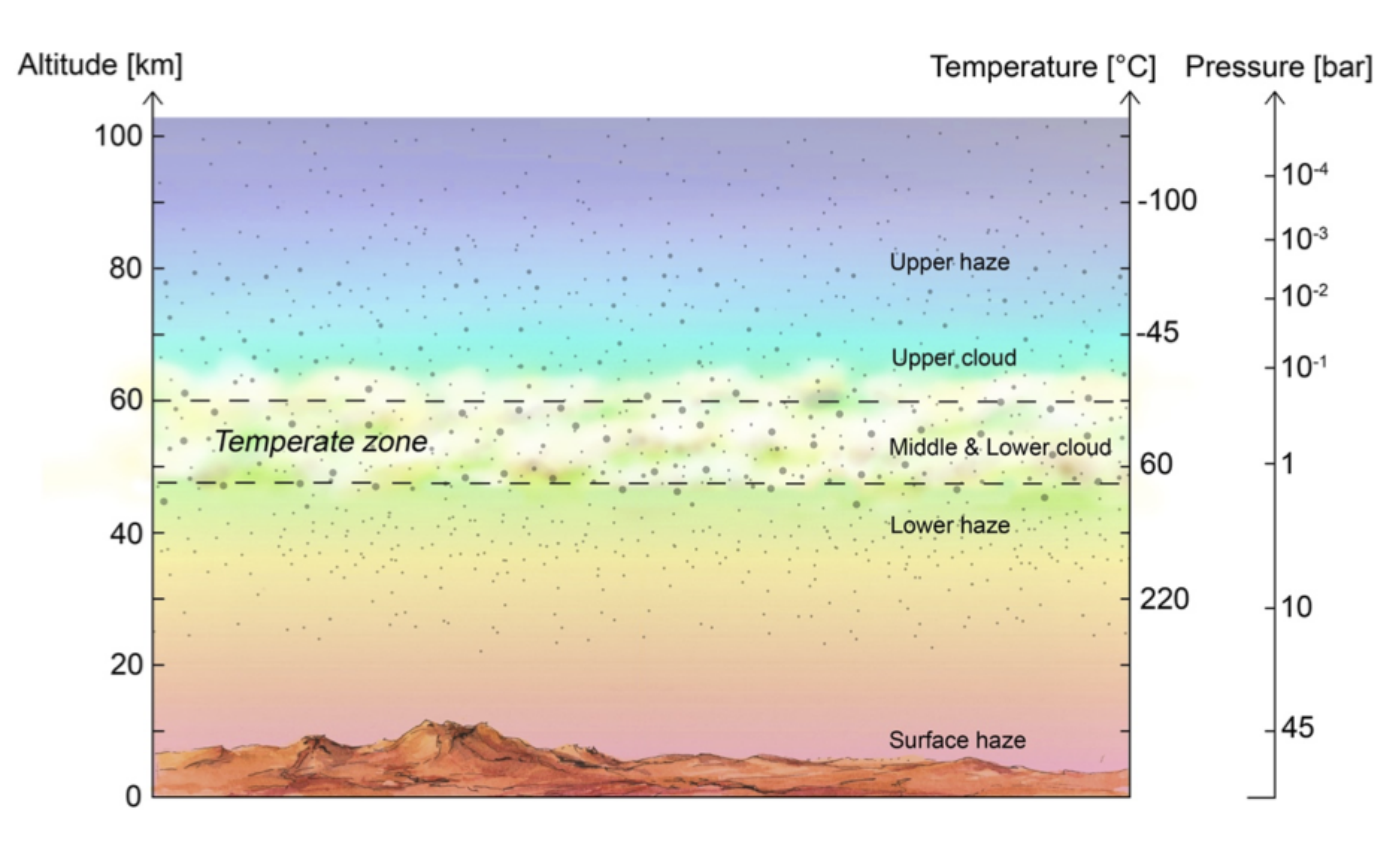}
\label{fig:atmosphere}
\caption{The atmospheric structure of Venus, showing how temperature and pressure vary with height. The cloud layer at around 55km altitude has temperature and atmospheric pressure levels that are comparable to those on the surface of the Earth. \cite{ss21}}
\end{figure}

\subsection{Venus in the past}

While the surface of Venus is undoubtedly hostile today, this might not have been the case in the distant past. If the surface of Venus was warm and wet - in the sense that liquid water could flow on the surface - then it is possible that life might have emerged there. When conditions later become increasingly hostile to life on the surface it might have evolved to seek sanctuary in clouds, the last habitable ecological niche on the planet. But is there any evidence to suggest that Venus was ever less hostile than it is today?

Sadly, we do not have direct access to information about the conditions on Venus billions of years ago, but there are hints that suggest that Venus may once have had much more water on its surface, and in its atmosphere, than it does today. Principal among this evidence is the deuterium to hydrogen ratio (D/H ratio). Venus currently has a D/H ratio that is 150$\pm$30 times that of the Earth \cite{d82} \footnote{Earth's D/H ratio is $\sim$ 1.6 $\times$ 10$^{-4}$}. This suggests that Venus has lost a substantial fraction of its water through the escape of hydrogen, which is favoured over the escape of deuterium since the latter has a higher mass. Hydrogen is driven off Venus through interactions with the solar wind which can directly interact with the upper layers of its atmosphere since, unlike the Earth, Venus lacks a protective magnetic field. However, it is possible that Venus may have retained a magnetic field, and thus the possibility of liquid water oceans, for several billion years after its formation \cite{gi22}.

Computer simulations have been used to asses whether a warm wet Venus in the first billion years of the Solar System would have a stable climate that could be conducive to the emergence of life \cite{w16}. While the conclusions of this work are still not agreed - some suggest that even a wet Venus would never have been able to condense its water into oceans, leading to a so-called `steam Earth' scenario \cite{tu21} - it is intriguing to consider the possibility that Venus may in fact have been the first habitable planet in the Solar System. A warm wet Venus (see Figure 4) might have remained habitable until as recently as about 700 million years ago, allowing substantial time for life to evolve and propagate across its surface given that life seems to have emerged on Earth somewhere between 3.7 and 4.3 billion years ago \cite{d17}. If this is correct, it may only be in the most recent 15\% of the age of the Solar System that massive volcanic activity on Venus established a runaway greenhouse effect on the planet, leading to water stripping and the hot, dry, hostile surface that we see today.

\begin{figure}
\includegraphics[width = 10cm]{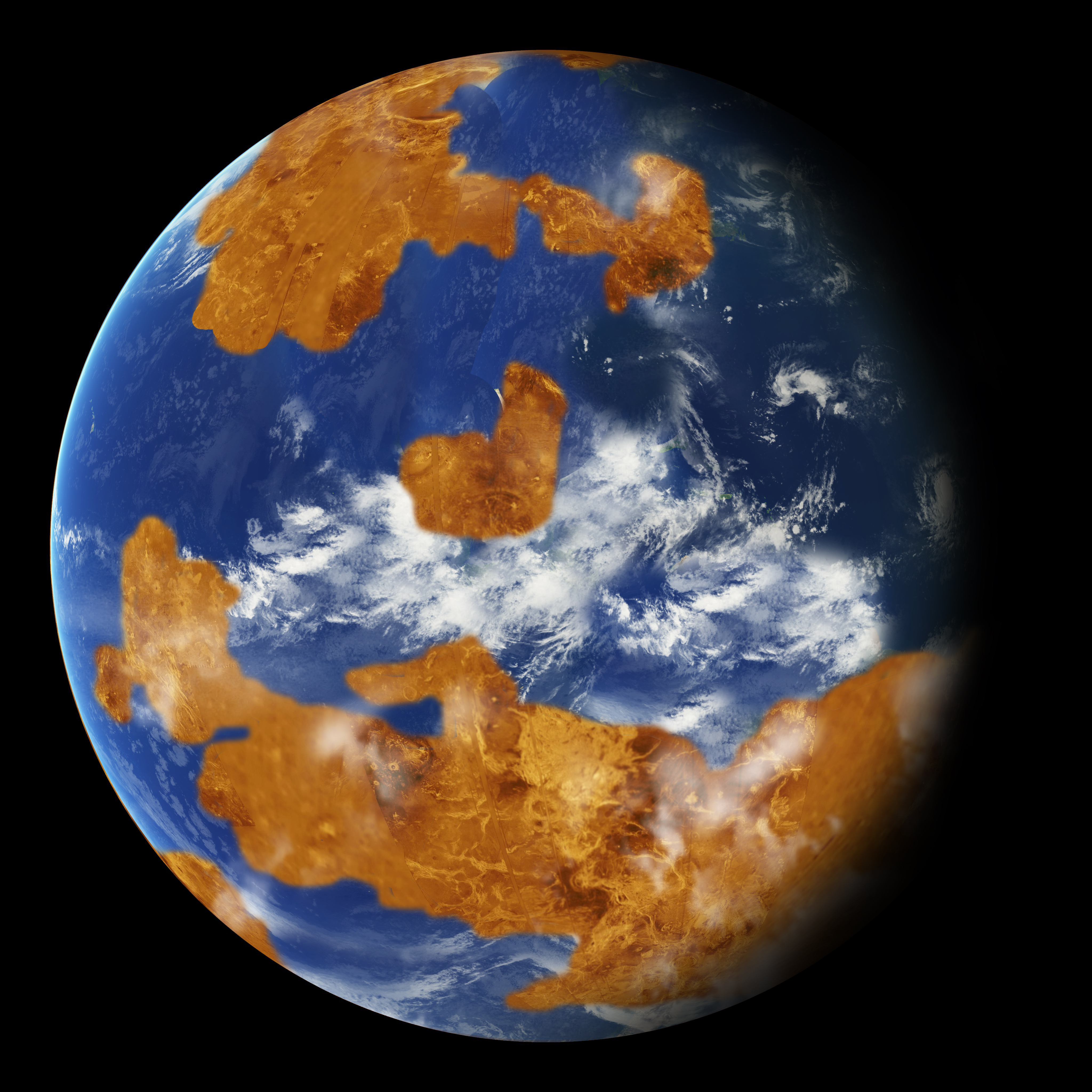}
\label{fig:wet_venus}
\caption{An artist conception of what a warm, wet Venus might have looked like during earlier stages in the evolution of the Solar System. Credit: NASA}
\end{figure}

\section{Atmospheric mysteries}

As well as uncertainty about the role of water in Venus' past, there are also some significant mysteries about the planet's atmosphere today even before we get to the detection of phosphine. Long-standing issues include (\cite{b21} and references therein):
\begin{itemize}

\item The variation of the abundance of water vapour and SO$_2$ with altitude in and above the cloud layers \cite{k06}. Water, H$_2$O, persists throughout the atmosphere but SO$_2$ levels drop from parts per million abundances below the clouds to parts per billion above. This is not what is expected given that both gases are thought to be released by volcanism  at the surface and to be well mixed throughout the atmosphere until both are destroyed by solar UV at altitudes of 70 km or higher. The apparent depletion of SO$_2$ in the clouds is currently not understood.

\item Oxygen, O$_2$, is present in the clouds of Venus where it was detected by gas chromatographs on board Pioneer Venus Probe \cite{o80} and the Venera 13 and 14 descent modules \cite{m82}. There is currently no known process by which oxygen can be formed in the cloud layers so its origin is something of a mystery. 

\item Observations of the clouds of Venus in the ultraviolet reveal complex spatial and temporal changes in absorption and reflection \cite{l18}. This is in contrast to observations in the optical and near-infrared where the clouds of Venus appear nearly featureless on the dayside (see eg. Figure \ref{fig:venus}). Significant cloud contrasts are only seen in reflected sunlight at wavelengths shorter than about 400 nm, and at near-infrared wavelengths (1.7 to 2.4 $\mu$m) in emission on the nightside. These variations in ultraviolet absorption were first observed in photographic observations in 1928 \cite{r28},  but, despite all the ground-based observations, spacecraft observations from orbit, and descent probes sampling the atmosphere, the chemical and physical origin of the absorber responsible remains unknown. 

\item The clouds of Venus contain a variety of constituents. Based on size analysis from the Pioneer Venus Probe particle size spectrometer \cite{k80} they can be divided up into three particle sizes. These correspond to aerosols, of size $\sim$0.4 $\mu$m, droplets of size $\sim$2 $\mu$m, and larger particles of size $\sim$7 $\mu$m. The larger particles are present only in the middle and lower cloud layers, at altitudes from 47.5 to 56.5 km above the surface. The nature of these largest particles, which may have a substantial solid component, and be non-spherical in shape, is currently unclear. 

\item Recent reanalysis of the chemistry of Venus' atmosphere based on measurements by mass spectrometers on descent probes suggest the possibility of chemical disequilibrium in the middle cloud layers \cite{m21}. This result is based on the presence of several species in the mass spectrometer data, including hydrogen sulphide, nitrous, nitric \& hydrochloric acids, carbon monoxide, ethane, and hydrogen cyanide as well as phosphine (this reanalysis was conducted after the detection of phosphine by ground based observations, of which more later) and possibly ammonia. This chemical mix indicates that reducing chemistry is taking place in the clouds. If so, then the processes behind this activity would be out of equilibrium with the oxidising chemistry of Venus. In the context of the search for life elsewhere, chemical disequilibrium is a potential biosignature, making these results very interesting. More recently still, it seems that ammonia, NH$_3$, may have been independently detected in Venus' atmosphere by ground based observations (Greaves et al., private communication), confirming the tentative results from the Pioneer Venus Probe mass spectrometer reanalysis, and adding an extra piece of evidence in favour of chemical disequilibrium in the clouds of Venus.

\end{itemize}

These problems with our understanding of the atmosphere of Venus, and specifically its clouds, make the case for renewed interest in the planet as a possible astrobiology target. Even if biological processes do not provide the explanation for these poorly understood phenomena, it is clear that there are chemical and physical processes underway in the clouds of Venus that we do not currently understand. Further observations to explore the chemistry of Venus and its atmosphere are thus needed. It is in this context that a team of astronomers proposed to look for phosphine, PH$_3$, on Venus\footnote{The author of the current paper was part of this team and is part of the ongoing work to study phosphine on Venus.}.

\section{The Search for Phosphine}

Phosphine, as we have seen above, has been proposed as a potential biosignature gas which might be present in significant quantities on inhabited planets orbiting other stars \cite{ss20}. Current observational facilities, however, are not yet able to make phosphine observations of terrestrial exoplanets. While we know that phosphine is present in the Earth's atmosphere in small amounts, thanks to industrial processes and anaerobic organisms, there were no limits on the amount of phosphine present on other Solar System terrestrial planets. Mercury has essentially no atmosphere so is an inappropriate target. Mars has a very thin atmosphere so is unlikely to have much phosphine even if there is biological activity underway there, and any phosphine present would be rapidly destroyed by solar UV radiation. This leaves Venus as the only reasonable terrestrial planet in the Solar System where some test observations in search of phosphine might be conducted.

Therefore in 2016 a team of astronomers led by Prof Jane Greaves put together a proposal to the James Clerk Maxwell Telescope (JCMT) to conduct test observations of Venus' atmosphere to look for absorption from the J=1-0 rotational transition of phosphine, which would produce an absorption line at a wavelength of 1.123 mm ($\sim$ 267 GHz). There are other transitions of phosphine at other wavelengths, notably in the far-IR and in the mid-IR, but this particular transition has some advantages, Firstly, observations can be conducted from the ground. Observations of the next highest rotational transition, J=2-1, would require observations from the stratosphere (see below). Mid-IR observations of other transitions can be conducted from the ground, of which more later, but the Greaves team did not have easy access to the necessary mid-IR facilities.

The initial idea for the observations was to acquire a few hours of data to better understand the observational issues with looking for a weak absorption line against a very bright continuum source, Venus, with the eventual intent to propose a longer series of observations to set a stringent upper limit, since phosphine was not expected to be found. That is not, however, how things turned out.

\subsection{JCMT Observations}

The JCMT is a 15 m diameter mm/submm telescope on the mountain Mauna Kea on the Big Island of Hawaii, at an altitude of about 4000 m. Mauna Kea is an ideal site for mm/submm observations since it is both high and dry, and so avoids much of the water vapour in the atmosphere of the Earth that would otherwise absorb and contaminate observations at these wavelengths. It is equipped with an array of instruments that operate both as continuum imagers and as high resolution spectrometers. For the first set of JCMT phosphine observations \cite{g21} an instrument called Receiver A3 (RxA3) was used. RxA3 was one of the early instruments used on the JCMT and was delivered to the telescope in 1998. It was retired not long after the phosphine detection observations reported in \cite{g21}.

Like many mm/submm spectroscopy receivers, RxA3 uses a heterodyne approach, whereby the incoming astronomical signal, in this case at frequencies of around 267 GHz, is multiplied by a pure sine wave signal at a nearby frequency, the so-called local oscillator frequency, using a device called a mixer. This  results in the production of a signal at a frequency that corresponds to the difference in frequencies between the received signal and the local oscillator frequency, and allows signals at frequencies outside the frequency range of interest to be removed. This lower frequency signal, at what is called the intermediate frequency, or IF,  is then measured and dealt with by later stages of processing. The technology used in most mm/submm receivers relies on SIS mixers (superconductor-insulator-superconductor) to mix the astronomical and local oscillator frequencies. For more information on how these operate, and on much else in radio astronomy, see \cite{w13}.

The IF signal from RxA3 is then processed by the Auto-Correlation Spectral Imaging System (ACSIS - this system is used by all spectral receivers at the JCMT, including both RxA3 and its replacement '\=U'\=u). This digitises the input IF signal, calculates the autocorrelation of the signal with itself - essentially multiplying the signal by a time delayed version of itself - and then calculates the Fourier transform of the autocorrelated signal. According to the Wiener-Khinchin theorem \cite{w13}, the autocorrelation of a signal is the Fourier transform of the signal's power spectral density ie. the amount power received as a function of frequency. Fourier transforming the autocorrelation of the IF signal thus gives us what we want - the spectrum of the source in the frequency range of interest.

\begin{figure}
\includegraphics [width = 14cm] {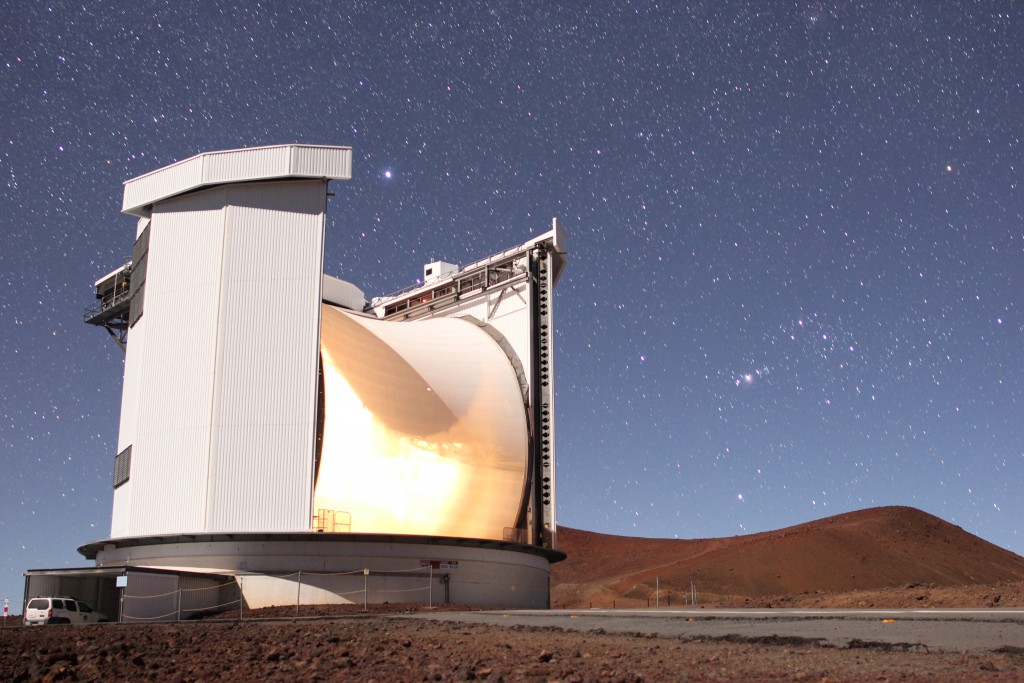}
\caption{The James Clerk Maxwell Telescope, a 15 m diameter mm/submm telescope on Mauna Kea in Hawaii, currently operated by the East Asian Observatory. The 15 m primary mirror is protected from wind during observations by a large gortex screen which is why it cannot be seen directly even when the telescope is taking observations, as in this picture. Credit: William Montgomerie/EAO/JCMT.}
\label{fig:jcmt}
\end{figure}

\begin{figure}
\includegraphics[width=14cm]{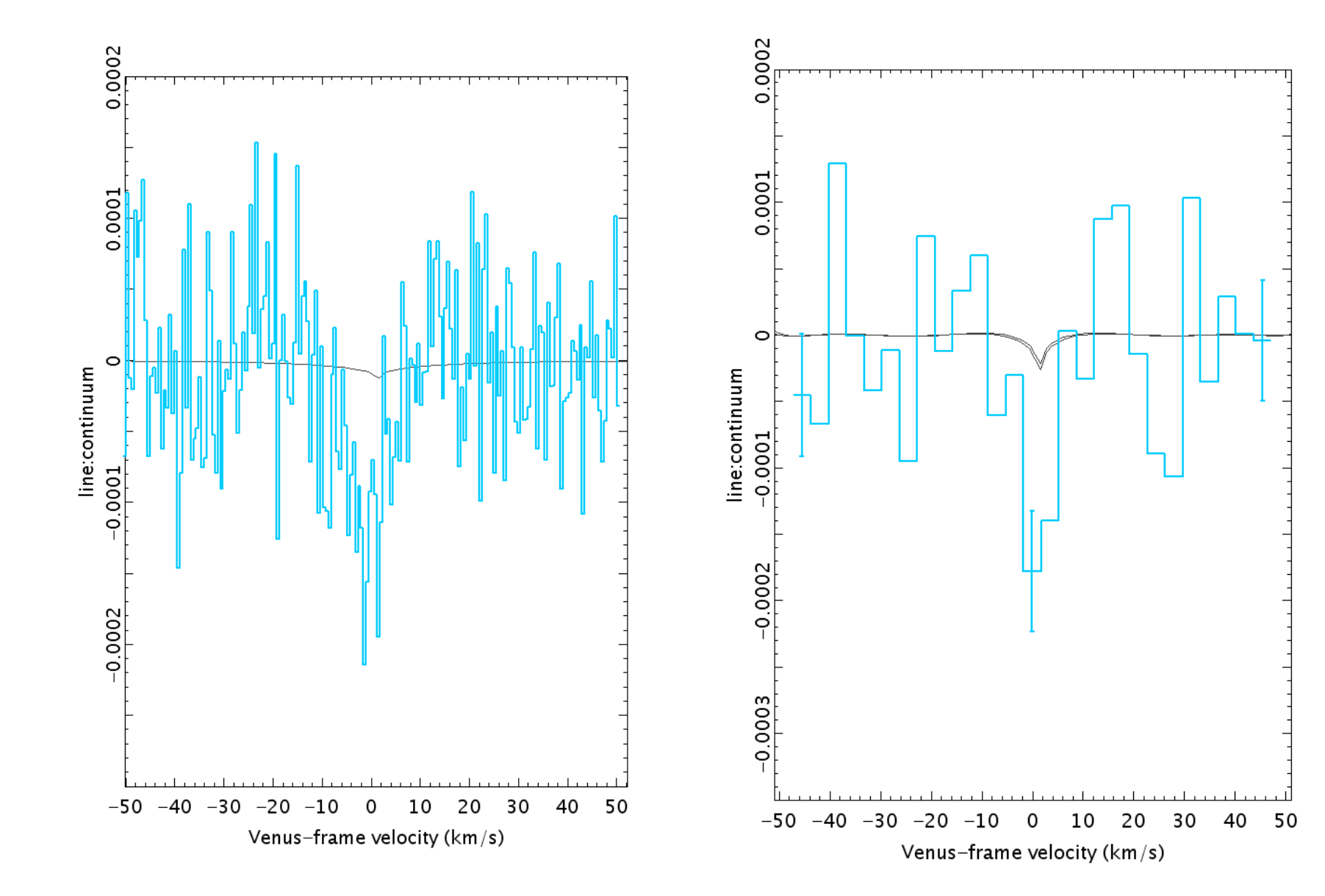}
\caption{The Phosphine 1.123mm J=1-0 line as detected by ALMA (left) \cite{g22} and JCMT with RxA3 (right) \cite{g21}. The black lines indicate the level of SO$_2$ absorption derived from simultaneous (ALMA) and near-simultaneous (JCMT) observations. As can be seen the PH$_3$ detections are clear and the SO$_2$ contamination is minimal. These spectra are continuum subtracted, so zero on the y-axis represents the continuum level. We use the standard astrophysics approach for presenting high resolution spectra in this Figure, where the spectrum is centred on the line of interest at zero velocity and frequencies are indicated by the doppler velocity in km/s needed to shift from this central value.}
\label{fig:phosphine}
\end{figure}

Venus was observed by the JCMT using RxA3 in search of phosphine on five mornings in June 2017. These dates were chosen so that Venus appeared large enough to fill the telescope beam, minimising any effects due to errors in pointing the telescope. Venus is a strong continuum emitter at millimetre wavelengths, so phosphine would be detected as a weak absorption line against this strong continuum. This strong continuum, however, leads to a number of problems with the quality of the data. A number of effects, including reflections from the floor or roof of the telescope dome, or in the receiver cabin itself, entering the beam, lead to strong, time varying baselines in the output spectra. These have to be detected and removed. For the initial JCMT detection of phosphine \cite{g21} these effects were removed by the usual method of fitting polynomial functions to the data, excluding the region of the spectrum where phosphine might lie. Once this process was applied to each of the 140 spectra that made up the observations, and despite the original assumption that only an upper limit would be found, an absorption line ascribed to phosphine was detected, corresponding to an abundance of about 20 to 25 parts per billion (ppb). The JCMT spectrum of phosphine can be seen on the right hand side in Figure \ref{fig:phosphine}.

\subsection{ALMA Observations}

\begin{figure}
\includegraphics [width=14cm] {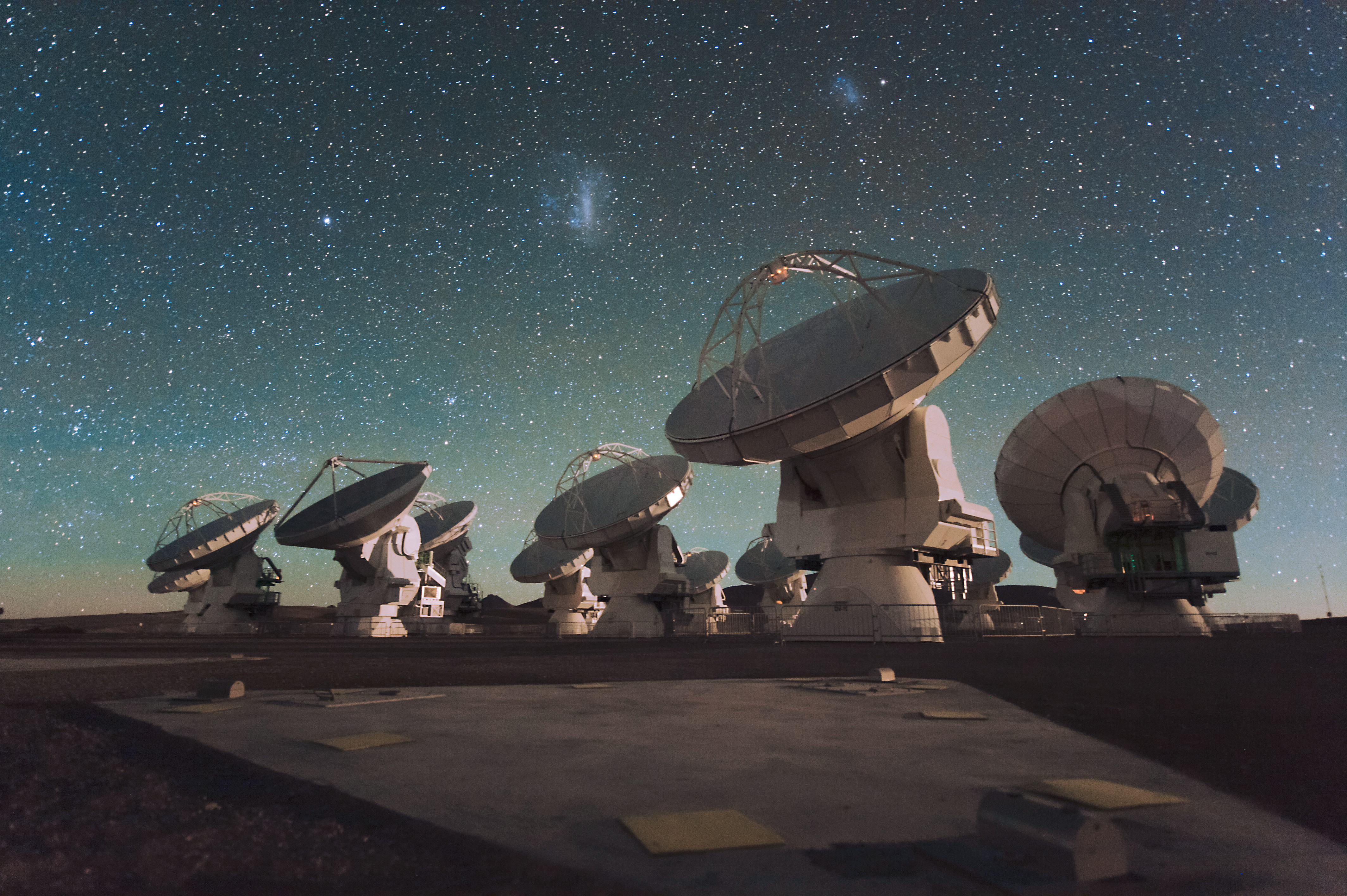}
\caption{Some of the 64 antennae that make up the ALMA telescope. Credit: ESO/C. Malin.}
\label{fig:alma}
\end{figure}

Following the rather surprising detection of phosphine at the JCMT some further observations in search of independent confirmation of this discovery were needed. To this end, observing time was granted on the Atacama Large Millimetre/Submillimetre Array (ALMA) in March 2019. Despite operating at similar mm/submm wavelengths, ALMA is a rather different telescope to the JCMT because it is an interferometer. It is made up of 66 separate antennae, mostly 12m in diameter, the signals of which are combined together to produce the final results. 43 of the 12 m antennae were used for the Venus phosphine observations.

While the signals received by each ALMA antenna are dealt with in a manner similar to RxA3 on the JCMT, using a heterodyne SIS mixer and local oscillator in the receiver, these signals are then cross-correlated pairwise with those from each of the other antennae in the array (where each pair of antennae forms a `baseline') to produce an interferometric map of the target. Interferometry allows angular resolutions to be achieved that correspond to a telescope whose diameter equals the longest baseline separating individual antennae. 

The cross-correlation of signals detected by each pair of antennae produces a series of `visibilities' which are a measure of the two-dimensional Fourier transform of the sky distribution of brightness. The visibilities at each observed frequency are then Fourier transformed to produce a series of images at successive frequencies, i.e. a spectral cube.  However, since only a finite number of antennae pairs are available, even for an array with as many antennae as ALMA,  the Fourier plane is like a telescope with lots of holes. Various methods are used in a process called `cleaning' to derive the actual image from the limited sampling in the Fourier plane. For more information on interferometry see \cite{w13} or the ALMA Primer\footnote{https://almascience.eso.org/documents-and-tools/cycle9/alma-science-primer}.

Processing interferometric data involves different challenges to those encountered at the JCMT.
For example, the angular size of Venus was so great that even the shortest ALMA baselines could not provide good images on the scale of the whole disc and the imperfect sampling led to strong ripples so the data from the affected short baselines, all less than 33 m in length, were removed.
There were also strong spectral ripples on some parts of the planet, such as the poles,
which had to be excluded from further analysis otherwise they would add noise to
the spectra, reducing the sensitivity of the final results.  
Further analysis of the
ALMA data processing also found some errors in the standard reduction script used, see Section 6.3.1, which 
improved on the initial detection. The end result of the ALMA observations once all these various effects are taken
into account is shown in Figure 6 - a good detection of phosphine absorption at a
level of $\sim$20 ppb that matches what was seen by the JCMT but with somewhat higher signal-to-noise.

%

\subsection{The Detection of Phosphine from the Ground}

The detection of phosphine in the atmosphere of Venus was, to say the least, a surprise. The observations from JCMT and ALMA thus prompted a considerable amount of debate and further observations using other facilities. In this section we look at these various discussions, their conclusions, and counter-arguments to the suggestion that phosphine has not been detected or that whatever has been detected was not phosphine.

\subsubsection{Reanalysis of the ALMA Data}

One of the first responses to the Greaves et al. detection paper, \cite{g21}, was a reanalysis of
the ALMA data by a separate group \cite{v21}.  This analysis did not reproduce the phosphine detection
of \cite{g21}, and instead found an upper limit to the phosphine abundance of about 1 ppb. They identified some processes used in the standard ALMA calibration scripts
which were not adequate for a very bright, time-varying, beam-filling target or indeed for the correspondingly bright calibrator sources used. This led to reprocessing of the raw data by the ALMA observatory and European Southern Observatory (ESO) staff (independently of any of the research groups), who provided new scripts taking these and additional problems into account. The reprocessing simplified the basic removal of instrumental bandpass ripples using the moon Callisto as a calibrator, and also avoided the chance of spectral averaging producing sharp edges which could mimic an absorption line.  The new scripts also accounted for the non-linear instrumental response to the high intensity of Venus (the brightest source in the sky after the Sun at these wavelengths) and its large angular size, although, since this exceeds the extent of accurate models of the response of individual ALMA dishes, this is thought to be a source of residual error.

Greaves et al. responded to this reanalysis \cite{g21b, g22} by employing the improved observatory scripts and updating their own processing, using three different independent methods to obtain final images and spectra. The first step after observatory calibration is to remove the shortest baselines as explained above, and then to make a simple, linear spectral fit to the visibility data to remove the contribution of Venus. Next, residual spectral ripples can be corrected either in the visibility data or after Fourier transforming to make an image cube, and before or after cleaning.  Spectra were extracted over different portions of the planet; small residual errors meant that only those spectra extracted from regions symmetric about the planet centre were considered reliable. A range of parameters allowed the  continued  recovery of a phosphine signal using all the updated methods, optimised at 7.7$\sigma$ significance by excluding the planetary poles \cite{g22}. They attributed the non-recovery of the phosphine signal by \cite{v21} as a result of including baselines shorter than 33 m in most of their analyses, as well as including parts of the image of the planet that had significant spectral artefacts that raise the noise in the final combined spectrum. They concluded that the phosphine detection in the ALMA data remained robust.

%

\subsubsection{Was it a real line?}

A common feature of both the original ALMA and JCMT data analyses in \cite{g21} was the use of fairly high order polynomials to allow the removal of varying baselines. In doing this, it is necessary to mask out the region of the spectrum around a suspected line otherwise the polynomial fitting method might fit and remove a real line, mistaking it for a small scale baseline ripple. Several authors suggested that this process can instead lead to the creation of fake lines, and that this was in fact the origin of the claimed phosphine detection \cite{v21, s20, th21}. There are two counter-arguments to this suggestion that the detection is essentially a statistical false positive. 

The argument that the claimed phosphine detection is a false positive is that when you take the ripple-contaminated spectrum, block out a portion of it where there might be a line, and use a sixth or higher order polynomial to fit the baseline, then some noise spikes or contaminating ripples in the blocked out section may end up looking like a line. This is in fact correct, and blind searches for line candidates at random locations in the spectrum would indeed suffer from this effect, significantly reducing confidence that any detections are real. However, the detection of phosphine in \cite{g21} did not solely rely on measuring the depth of an absorption line at a random position. Instead, it also relied on the wavelength of the line seen coinciding with that of the line being searched for, phosphine. This significantly reduces the chance of a noise spike or residual masquerading as a phosphine detection. Analysis in \cite{g21c} shows that adding the additional constraint that a fake line must be at a specific frequency reduces the chance of a false positive for line detection to $< 1.5\%$.


Furthermore, if the line was in fact a false positive then there would be no reason for any such noise-generated feature to lie at exactly the same frequency in both the ALMA and JCMT data. As pointed out in \cite{g21}, the only feature at matching wavelengths in both the ALMA and JCMT data lies at the expected frequency of phosphine. This further bolsters our confidence that the detected phosphine line is real, and not a statistical artefact resulting from the data processing approach.


\subsubsection{Is it really phosphine?}

The foregoing analysis suggests that the line discovered is in fact real and not a statistical false positive. However, can we be sure that it is in fact phosphine and not some other molecular species that happens to have an absorption feature at a similar frequency? Sulphur dioxide, SO$_2$, a known constituent of Venus' atmosphere, has a transition due to the (J = 30$_{9,21} - 31_{8,24}$) transition at 266.943329 GHz, a frequency shift from the PH$_3$ J = 1-0 line at 266.944513 GHz that corresponds to a velocity difference of just 1.3 km/s. The possibility that the claimed phosphine line is actually a misidentification of this SO$_2$ line was first suggested by \cite{v21} and has been further explored by others \cite{a21, l21}. While they have concluded that SO$_2$ contamination or misidentification is a possibility, a number of problems with this interpretation have been pointed out by \cite{g22}. Firstly, while the line centres of PH$_3$ J=1-0 and SO$_2$ J = 30$_{9,21} - 31_{8,24}$ are close, they are still 1.3 km/s apart, leading to a $\sim$ 3 $\sigma$ discrepancy between the measured line centre and that expected for the SO$_2$ line. Furthermore, simultaneous (in the case of ALMA) and near-simultaneous (in the case of JCMT) observations of a different and stronger SO$_2$ line \cite{g22} provide predictions of the relative strength of the SO$_2$ transition that might contaminate the phosphine line. They find that the level of contamination of the phosphine line by SO$_2$ is $\sim 10\%$ for the JCMT data and $< 2\%$ for the ALMA data. This level of contamination by SO$_2$ is shown as a black line in Figure \ref{fig:phosphine}. On this basis it seems likely that the detected line is indeed phosphine, and that any contamination by the neighbouring SO$_2$ line is insignificant.

\subsubsection{Other Observations}

The phosphine J=1-0 line at 1.123 mm is not the only line of this molecule. However, many of the other transitions are at wavelengths that are more difficult to observe from the ground. Nevertheless, observations have been attempted of other lines in search of independent confirmation of the presence of phosphine.

The first of these used archival data from the TEXES (Texas Echelon Cross Echelle Spectrograph) instrument, a 5 to 25$\mu$m high resolution mid-infrared spectrometer, on the NASA Infrared Telescope Facility (IRTF) on Mauna Kea in Hawaii \cite{e20}. These observations were part of a long term project to monitor SO$_2$ and H$_2$O in the cloud tops of Venus, and involved observations at a range of frequencies. One of these datasets, obtained in March 2015, fortuitously included a range of wavelengths where there are some relatively strong phosphine transitions, at a wavelength around 10.471 $\mu$m (corresponding to a frequency of 28.65 THz). No phosphine absorption is detected, indicting an upper limit of about 5 ppb, which is substantially lower than the claimed millimetre wave phosphine detection.

Further infrared data, this time from the Venus Express spacecraft, were analysed, looking for absorption from phosphine lines at wavelengths around 4.125 $\mu$m above the cloud layers \cite{t21}. This data was taken at various times from June 2006 to December 2014, and measured absorption against the light of the Sun as it rises or sets, rather than against the emission of Venus itself. This means that only a small part of the atmosphere is studied rather than the entire planetary disk as is the case, for example, for the JCMT or TEXES observations. These Venus Express observations also failed to find any phosphine absorption, setting limits on its abundance of 0.2 to 20 ppb depending on the specific observations and the assumed altitude of the absorption, ranging from 60 to 95 km.

\begin{figure}
\includegraphics [width=12cm]{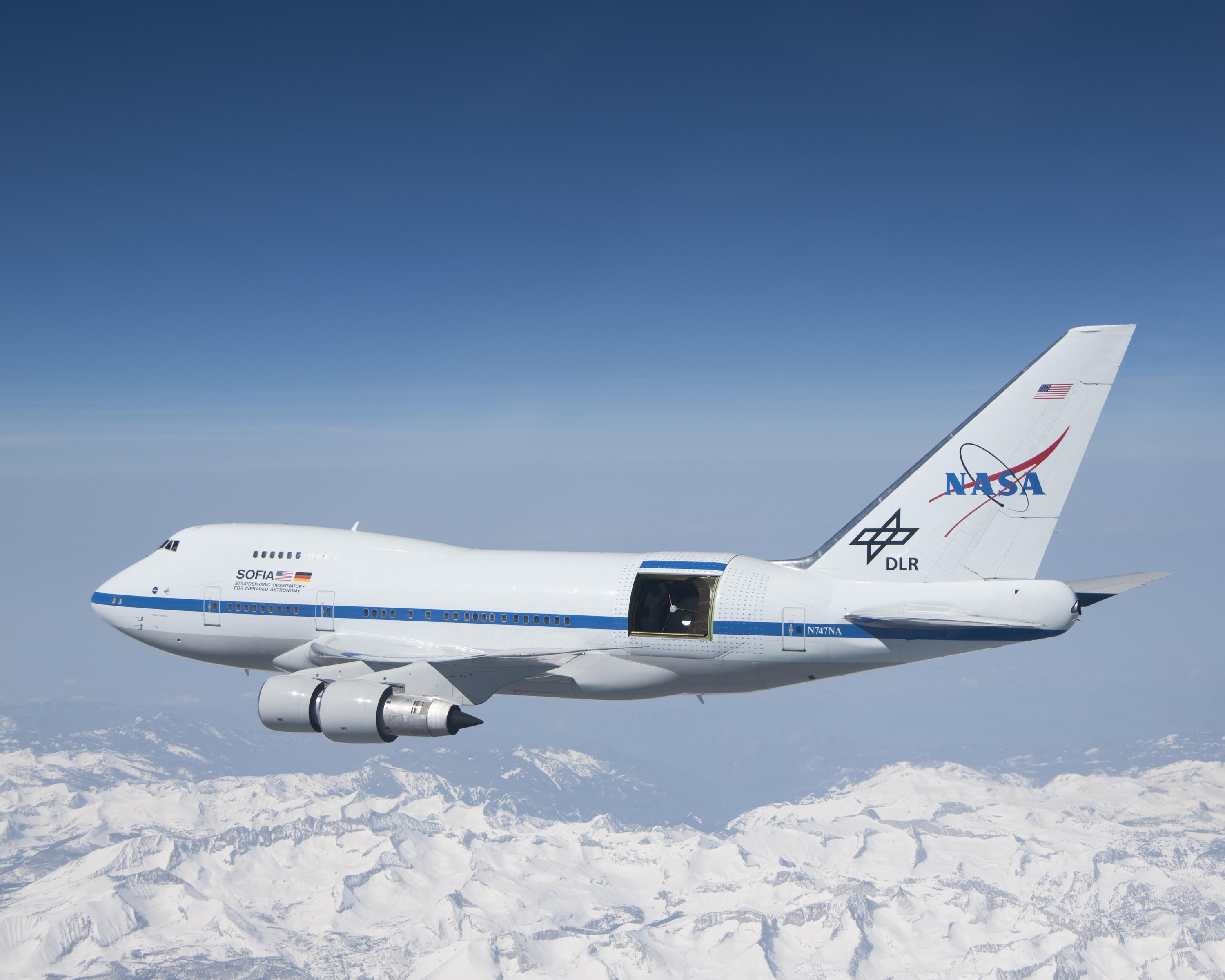}
\label{fig:sofia}
\caption{The SOFIA observatory, which consists of a 2.5m telescope and instruments mounted inside a 747 jumbo jet, and capable of flying above much of the far-IR absorption in the Earth's atmosphere. Credit: NASA/DLR}
\end{figure}

A third approach to confirm the detection of phosphine is to search for absorption lines in the far-infrared, at frequencies around 534 and 1067 GHz \cite{c22}. While the Earth's atmosphere is completely opaque at these frequencies at sea level and even on tall mountains like Mauna Kea, the SOFIA observatory (Stratospheric Observatory For Infrared Astronomy) - essentially a 747 Jumbo Jet with a hole cut in the fuselage with a 2.5m telescope pointing out (see Figure 8) - can perform these observations since it flies at an altitude of about 13 km, above much of the water vapour that absorbs far-IR radiation in the Earth's atmosphere\footnote{Sadly such observations can no longer be performed since the SOFIA observatory was decommissioned and retired at the end of September 2022.}.

SOFIA observations of Venus in search of phosphine were carried out in November 2021 \cite{c22} using the GREAT (German REceiver At Terraherz frequencies) instrument, a receiver similar to the JCMT receivers but operating at much higher frequencies. Data reduction and analysis by the original authors failed to find any sign of phosphine, setting an upper limit of 0.8 ppb from the J=4-3 line and $\sim$ 2 ppb for the J=2-1 line. However, subsequent reanalysis of the SOFIA data found that the calibration stage that sets an absolute flux scale adds noise and artefacts to the resulting spectra. This calibration stage is not needed if we are only interested in the line-to-continuum ratio, as is the case when measuring an absorption line. By purely analysing the line-to-continuum ratios \cite{g23}, phosphine at a level of $\sim$1-2 ppb is found, averaged over altitudes from 75-110 km, with 6.5$\sigma$ significance.

These other observations in search of phosphine absorption using different approaches, whether from the ground or from Venus Express, have produced a number of conflicting results. They need to be carefully interpreted since the different wavelengths and observational approaches are in fact probing the presence of phosphine at different altitudes and times, as we shall see below. None has yet definitively disproved the original JCMT and ALMA results of \cite{g21}, and the SOFIA observations may in fact have provided some level of confirmation, depending on which analysis approach is used.

\subsubsection{{\em In Situ} Confirmation}

The ideal way to determine the presence and amount of phosphine in the atmosphere of Venus would be to send a space probe directly into the atmosphere equipped with instrumentation that can detect and measure the presence of the gas {\em in situ}. This would avoid all the difficulties of observing phosphine remotely, as well as all issues of interpretation. At this point, as we will see below, we are some years away from any such future mission. However, past missions to Venus did send probes into the planet's atmosphere. Principal among these, for our current purposes, is the Pioneer Venus Multiprobe (also known as Pioneer Venus 2 or Pioneer 13) \cite{d79} (see Figure 9) which, among many other instruments, sent a mass spectrometer into the atmosphere of Venus on its largest entry probe.

\begin{figure}
\includegraphics [width=12cm] {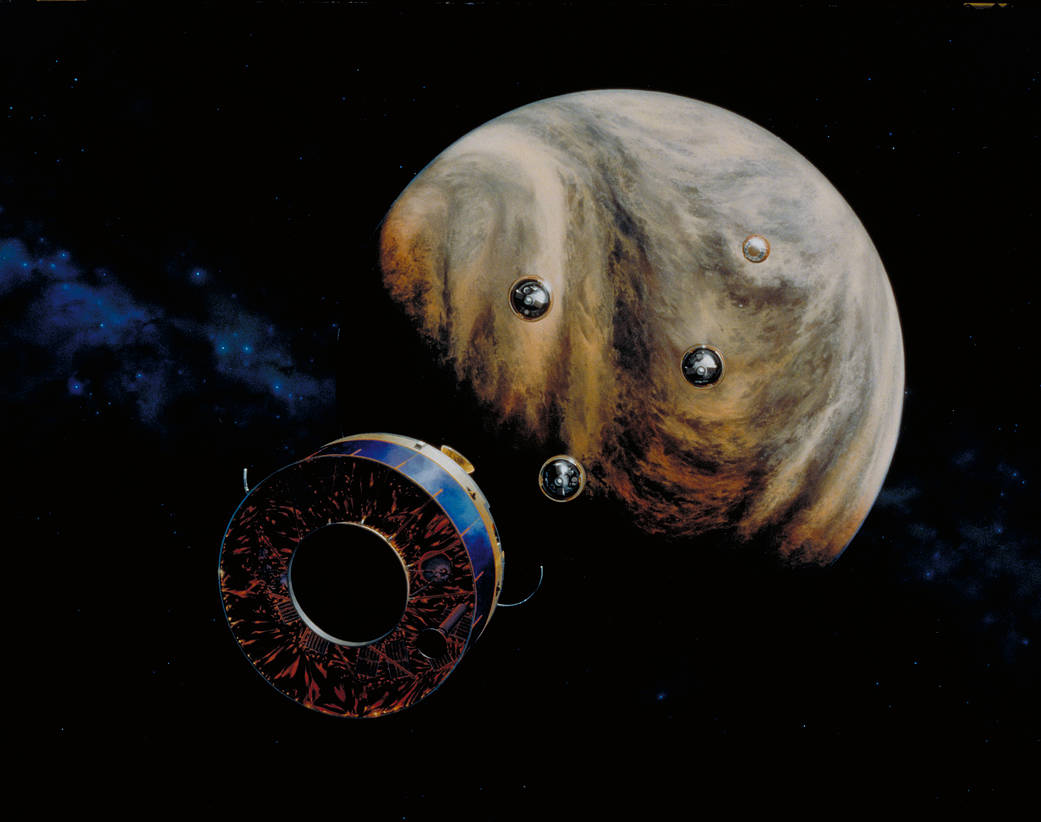}
\label{fig:pv_probe}
\caption{The Pioneer Venus Probe Spacecraft. The mission consisted of an orbiter and four probes that were sent into the atmosphere of Venus, seen here in artists conception. Credit: NASA}
\end{figure}

Data from the Pioneer Venus Large Probe's Neutral Mass Spectrometer (LNMS) were reanalysed in 2021 \cite{m21} subsequent to the announcement of the discovery of phosphine by \cite{g21}. This reanalysis of data taken during its descent into the atmosphere of Venus on 9 December 1978, was the first to look for trace or minor constituents of the atmosphere beyond methane and water. The LNMS takes gas in from the atmosphere through inlet tubes. Molecules in the gas are then ionised by an electron source, accelerated by an electric field and then passed through a magnetic field which deflects the ions by an amount that depends on their mass. These ions are subsequently detected, allowing their mass and abundance to be determined. For more information see \cite{h79}.

The data reanalysed in search of phosphine came from within the clouds, at an altitude of 51.3 km above the surface, part of the atmosphere that is largely inaccessible to ground or space based observations, but which is of critical importance to searches for possible life in the clouds, as this is where that life might actually live. The detailed analysis found evidence for phosphine at 0.1 to 2 parts per million (ppm) levels in the clouds themselves, a much higher abundance than is seen in the JCMT or ALMA observations. It also found evidence for other species such as nitrite, nitrate, nitrogen and possibly ammonia. Taken together these molecules indicate that unexpected chemical processes are underway in the clouds and suggest chemical disequilibrium. Whether this disequilibrium is due to biological or some other as-yet unknown chemical process is yet to be determined.

\subsection{Where and When is the Phosphine Seen?}

The forgoing sections, describing the various observations in search of phosphine in the atmosphere of Venus, can seem confusing and mutually contradictory. This is at least partly because observational constraints mean that they sample the atmosphere of Venus at different altitudes and, because they encompass datasets that span over 40 years, at different times. We already know that some species in Venus' atmosphere are highly variable - SO$_2$ levels, for example, can vary by large factors on timescales of both years and days at various altitudes \cite{v17} - so this may also apply to phosphine. Spatial variations across the disk of the planet are also possible, but this is difficult to assess for phosphine since many of the observations to date have been of the average phosphine level across the planetary disk. 

Of particular importance is the amount of phosphine in the atmosphere as a function of altitude. While the {\em in situ} observations of the LNMS and Venus Express have a clearly determined altitude, this is harder to extract for the observations from Earth. In principle, the effect of pressure broadening on the absorption lines can be used to determine the vertical abundance profile of an absorbing molecule. Pressure broadening of an absorption line occurs when the absorbing molecules interact collisionally with other molecules in the atmosphere. These interactions shorten the characteristic time of the absorption process, in accordance with Heisenberg's uncertainty principle, increasing the uncertainty of the absorption frequency and thus broadening the line (see eg. \cite{p81}). The overall effect is to make the line shape a Lorentzian function, which has much broader wings than the usually assumed Gaussian shape. The exact width of the Lorentzian line depends on the pressure, temperature and nature of the molecules that are interacting. The higher the pressure, the broader the wings, so a full analysis of the shape of the phosphine absorption line can reveal its vertical abundance profile in the atmosphere of Venus.

There are, however, a number of problems with a full pressure broadening analysis of the phosphine line seen in the atmosphere of Venus. Firstly, the pressure broadening coefficient for phosphine in CO$_2$, the dominant constituent of Venus' atmosphere, is not currently known. Analyses have so far used either a modification of the phosphine broadening coefficient in air \cite{e20} or have used the CO$_2$ pressure broadening coefficient for NH$_3$ as an analog to that of phosphine \cite{g21}. Secondly, and more significantly for the immediate understanding of phosphine in Venus, the data reduction techniques used to date to extract the absorption line remove any broad line wings as part of the process that removes baseline ripples. This just leaves narrow line cores, meaning that the observations are insensitive to any significantly broadened lines, and thus are only sensitive to phosphine at altitudes of 75 to 80 km. Most recently, an experimental data processing approach applied to new observations of Venus from the JCMT-Venus project (PI: D.L. Clements) seems to be able to recover the broad line wings of the J=1-0 phosphine line, suggesting an abundance of phosphine at the ppm level inside the clouds at an altitude of about 60 km, consistent with the high levels seen by the LNMS.

The other factor to consider is the timing of the observations. While we do not yet have enough observations to allow us to monitor any changes in the abundance of phosphine with time or in relation to other species such as HDO or SO$_2$, we can see if there are any correlations between the amounts of phosphine seen and the timing of the observations relative to the illumination of Venus' atmosphere by the Sun. This may well be an important factor since photolysis by sunlight is a significant destruction route for phosphine in the Earth's atmosphere \cite{ss20}. If we combine all the phosphine observations - detections and non-detections - together with information about whether the Sun is rising or setting on the atmosphere at the time of observation we perhaps begin to see a pattern (see Figure 10).

\begin{figure}
\includegraphics[width=14cm]{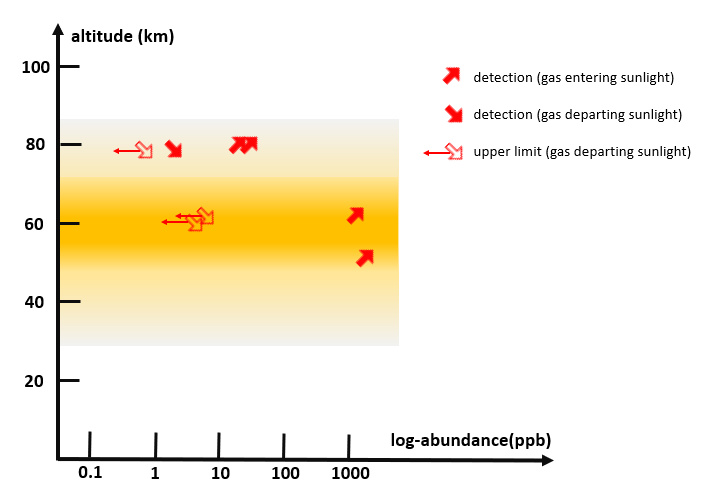}
\label{fig:sunlight}
\caption{The trend of phosphine abundance by altitude from the currently available data. Shading indicates cloud (orange, centred at $\sim$ 60 km) and haze (grey, centred at $\sim$ 80 km and 40 km) layers of Venus' atmosphere. Superposed symbols indicate candidate detections plus upper limits for phosphine abundance. Rising arrows indicate observations made where the atmosphere was rising into sunlight and falling arrows indicate observations made when the atmosphere was descending towards the nightside. Abundances are, from top: ~20, 25 ppb from J=1-0 data \cite{g21}; $\sim$1 ppb  or $<$ 0.8 ppb from J=4-3 data \cite{c22, g23}; $< 7$ ppb at 62 km from 4 $\mu$m spectra \cite{t21}; $<$ 5ppb at 60 km from 10 $\mu$m spectra \cite{e20};  $\sim$2 ppm at 60 km from initial JCMT-Venus processing; high ppb to 2 ppm at 51 km from Pioneer-Venus {\em in situ} sampling \cite{m21}. As can be seen all the significant detections of phosphine take place as the atmosphere is moving out of night and into sunlight, while the non-detections take place as the atmosphere is moving from sunlight into night. If sunlight destroys phosphine at high altitudes during daylight, as is the case on Earth, this would explain the apparent contradictions between some of the observations. From: Greaves et al. in prep, by permission.}
\end{figure}

\section{The (Im)Possible Origins of Phosphine on Venus}

The presence of phosphine in the atmosphere of Venus is a surprise since a compound of phosphorous with hydrogen should not naturally appear in the atmosphere of a planet, such as Venus, which has an oxidised atmosphere. On Earth, phosphine does not occur through normal chemical processes and is produced only by anaerobic life or through human industrial activity. While this is the expectation, Venus is a complex environment with a wide range of chemical and physical processes underway from the surface to the top of the atmosphere. A detailed analysis is thus necessary to see if there are any possible routes through which the levels of phosphine seen might occur through normal chemical processes. Such an analysis was conducted in \cite{b21b} where a wide range of chemical processes were examined to see if there is any potential source of phosphine in sufficient abundance to explain the observations with processes that we know are underway on the planet. The processes examined included gas reactions, geochemical reactions, photochemistry, volcanism (see also \cite{b22}), lightning and impactors. An example of the kind of chemical reaction network considered is shown in Figure 11, where the reaction rate and the destruction rates are compared. Only segments of this reaction network where the ratio of the production rate over the destruction rate is $\geq1$ can produce an accumulation of the relevant chemical. For phosphine to be produced in significant amounts the whole reaction network must have this ratio $\geq 1$ but, as can be seen, critical segments of the network have ratios orders of magnitudes less than this. More generally, it was found that the lifetime of free phosphine at various altitudes in the atmosphere of Venus ranged from $<$ 1 second to perhaps a century \cite{b22}, making it highly unlikely that a significant amount of phosphine can accumulate from any hypothetical source.

\begin{figure}
\includegraphics[width = 14cm]{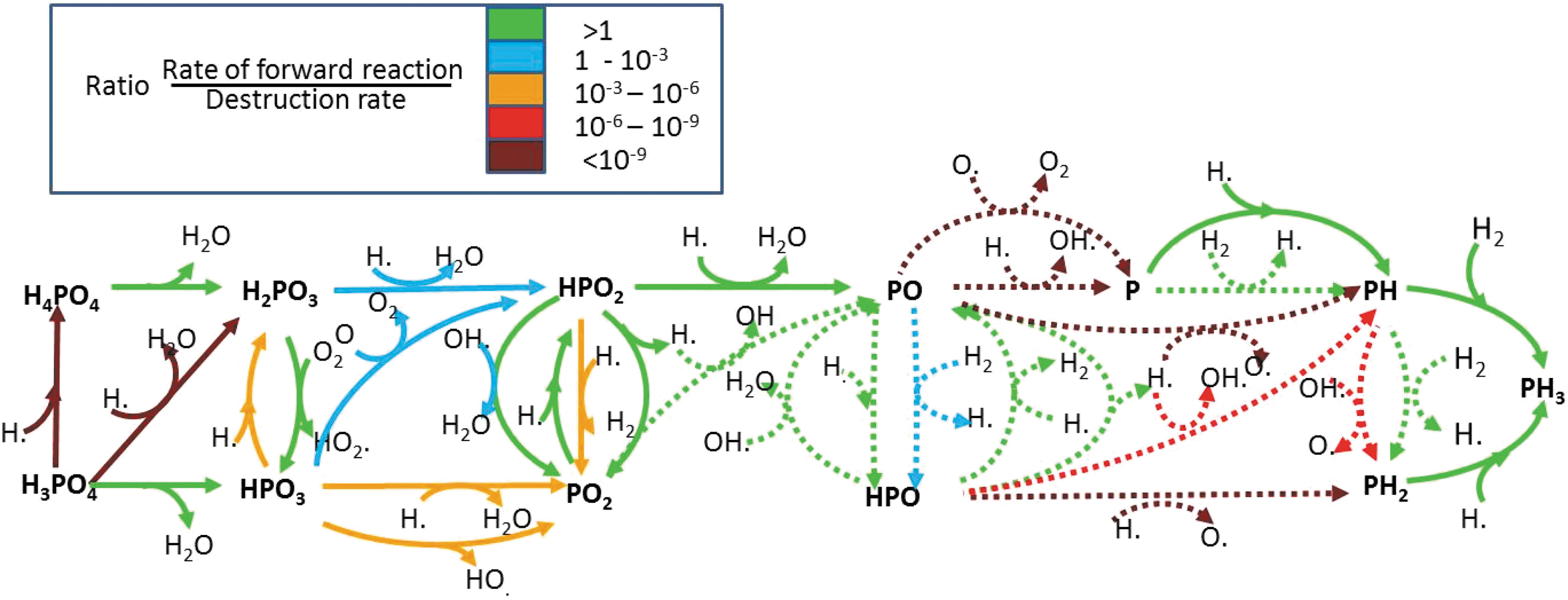}
\label{fig:chemistry}
\caption{Potential chemical pathways for the synthesis of phosphine in the atmosphere of Venus, and their derived production vs. destruction rate. There are  stages where, for all possible pathways, the rate of destruction of phosphine exceeds its formation by many orders of magnitude, as shown in red/purple. As can be seen, there is no route to produce phosphine by these processes that can account for the amounts observed. From \cite{b21b} where more details can be found.}
\end{figure}

The most obvious conclusion that can be drawn from this analysis is that we do not know how phosphine came to be in the atmosphere of Venus. There may be geochemical or photochemical processes that can produce it in sufficient amounts, but these are currently not known to us. The alternative, that, by analogy with Earth, phosphine is being produced by anaerobic biological processes, is another potential explanation. However, before we can make this particular leap, and claim that we have found evidence for life in the clouds of Venus, we must first exclude all other possible origins, and also explain how life is able to survive in the extremely acidic environment of Venusian cloud droplets. One possible solution to the latter problem is that ammonia, if present, is able to buffer the sulphuric acid in these droplets to some extent \cite{b21}. The possible detection of ammonia in the clouds of Venus by the LNMS \cite{m21} and in preliminary analysis of data from the Green Bank Telescope (Greaves et al., private communication), is thus rather interesting. 

\section{The Next Steps}

As has become clear in the previous section, studies of phosphine, and the search for life on Venus, are very much works in progress. While the current results are intriguing, there are no solid conclusions that can yet be determined. Much more work needs to be done, and it will be the work of many years before we can have a definitive answer to the question of whether there is life in the clouds of Venus. This will require not only observations from Earth, but also {\em in situ} probes and, ideally, missions that can return samples from Venus to Earth. In this section we look at some of the projects that are planned or already underway to improve our knowledge of the clouds of Venus.

\subsection{Earth-based Studies}

Observations from Earth were responsible for the first detections of phosphine, and these are continuing to both monitor phosphine and to search for other molecules that may have a bearing on the chemistry, or biochemistry, underway in the clouds of Venus.

The largest of the projects currently underway is JCMT-Venus (PI: D.L. Clements). This uses the  '\=U'\=u receiver, the replacement for RxA3, together with the ACSIS system to obtain whole disk spectra for Venus. The new receiver has a wider bandwidth than RxA3 so we can simultaneously observe phosphine, HD and SO$_2$, and search for other molecules such as SO and PO$_2$ which have spectral features in the band covered by  '\=U'\=u. By simultaneously monitoring phosphine, HDO and SO$_2$ we can see how these different species vary in relation to each other. This should provide indications as to the chemical processes behind the presence of phosphine. If, for example, phosphine is produced by reducing processes in the upper atmosphere, the proportions of reduced compounds, like phosphine, and oxidized compounds, like HDO and SO$_2$, will be anticorrelated. The JCMT-Venus project is a long term programme at the JCMT and has been awarded 200 hours of time over a period of three years. The visibility of Venus means that observations will be possible in three tranches, including Feb 2022, July 2023 and September 2023. The first of these observing campaigns has already taken place, with Venus observed over a period of 20 consecutive mornings. The data obtained already contains 140 times as much information as in the original JCMT observations, so is taking some time to process and analyse, especially since '\=U'\=u has its own difficulties dealing with the brightness of Venus and thus an interesting new set of baseline drifts and ripples to be removed. Nevertheless the analysis is well underway and initial results, some of which have been briefly discussed above, have already emerged, including further confirmation of the presence of phosphine. When complete, JCMT-Venus will provide a major new database of observations of Venus in the mm band, including phosphine and other important molecules which will provide significant new insights into the origin of phosphine. 

Further ALMA observations have yet to be approved, but these hold the promise of providing further information about the distribution of phosphine across the face of the planet. The original ALMA observations provided some hints that the distribution is not uniform, but a full map of the abundance of phosphine across the planetary disk could not be made because of excess ripples affecting the signal over significant portions of the disk. Additional observations in the mid-IR from the IRTF and elsewhere will also be helpful. Sadly, further observations with SOFIA are not possible since the observatory has been decommissioned.

Studies related to the search for potential signs of life on Venus are also underway that do not directly target phosphine. These include observations with the Green Bank Telescope (GBT) at radio wavelengths to look for ammonia (NH$_3$) in absorption. This is important since detection of ammonia would indicate the presence of another reduced molecule that should not be expected in the oxidised atmosphere of Venus. Ammonia is also important because of its buffering effect against the high acidity in the liquid droplets in Venus' clouds \cite{b21}. Analysis of archival data from the 1970s as well as an initial set of observations with the GBT suggest the presence of ammonia (Greaves et al., private communication), as do the {\em in situ} measurements of the LNMS, but more data is needed to confirm this.

Laboratory studies also have a role to play since they can validate and test the various assumptions that went into the analysis of \cite{b21}, and allow more accurate predictions for the formation and destruction of phosphine and other hydrogen-rich compounds in Venus-like conditions. Such studies are already being planned.

\subsection{Space-based studies}

Venus is also being studied from much closer quarters by space probes. These missions take many years to prepare and so have largely not been designed to examine the possibilities of unusual chemistry, or even life, in the clouds of Venus. Nevertheless, existing missions do have useful capabilities for these purposes and future missions are being planned that can respond to recent discoveries.

The Japanese mission AKATSUKI (see eg. Figure \ref{fig:venus}) is currently operating in orbit around Venus. While it does not have any instruments that are directly relevant to the search for phosphine, its UV imaging instruments are monitoring the unidentified UV absorber, the origin of which is one of the outstanding mysteries of the Venusian atmosphere. Comparing AKATSUKI's results with future data from the ground, especially any future imaging observations with ALMA, may be able to see if there is a link between the presence of phosphine and the presence of the UV absorber.

The next potentially important space mission to go to Venus, from the point of view of phosphine observations, is not in fact a specific mission to Venus, but the JUICE mission to the moons of Jupiter \cite{g13}. The JUICE spacecraft is scheduled to be launched in the second quarter of 2023. It will then perform a series of flybys of planets as gravity assists on its journey to Jupiter. Of particular interest here is the flyby of Venus in August 2025 where an observational campaign is possible. Of particular importance in the context of phosphine on Venus is the Submillimetre Wave Instrument (SWI) which will be able to observe higher J transitions of phosphine, including those observed from the Earth by SOFIA. Whether JUICE will be able to undertake an observing campaign at Venus will be up to the JUICE mission directors, and no decision will be made until after launch. 

In the 2030s three missions directly targeted at Venus are due to be launched. These include the European Space Agency's EnVISION mission \cite{gh21}, and NASA's VERITAS  (Venus Emissivity, Radio Science, InSAR, Topography, and Spectroscopy) \cite{sm21} and DAVINCI (Deep Atmosphere Venus Investigation of Noble gases, Chemistry, and Imaging) \cite{ga22} missions. VERITAS and EnVISION are primarily concerned with the surface and interior of Venus, studying the history and role of volcanism on the planet. While they will doubtless reveal much that is of interest, they are unlikely to have much to say about phosphine and the processes underway in the clouds of Venus unless they uncover volcanic activity vastly in excess of our current understanding \cite{b22}. DAVINCI, however, is a much more interesting prospect.

The goal of DAVINCI is to study the atmosphere of Venus. To do this its primary set of instruments are on board a descent stage that will fly through the clouds of Venus, sampling the atmosphere as it goes. It will be the first NASA mission to enter the atmosphere of Venus since Pioneer Venus Probe in 1978. Among the instruments on the descent stage is a mass spectrometer that will be able to significantly improve on the results of the LNMS. This will be able to detect phosphine and other trace gas species and see how their abundance changes with altitude and other conditions. Other instruments include a tuneable laser spectrometer which is able to measure even small amounts of specific gases. Altogether, the four instruments on the DAVINCI probe, combined with imagers on the orbiting mothership, will provide a vast improvement in our {\em in situ} knowledge of Venus' atmosphere. It will provide the ground truth against which observations of the planet from Earth can be compared.

National and international space agencies are not the only organisations looking to send probes to Venus. Private companies now have the capability to send missions to other planets independently of governments, and they are also interested in the possibility of life on Venus. One company in particular, Rocket Lab, is taking special interest in Venus and has set up a team to develop a series of Venus Life Finder (VLF) missions \cite{s22}. The first of these missions, which may launch as soon as mid-2023, is intended to look for organic molecules using an ultraviolet autofluorescence technique. Further missions are planned including a balloon borne laboratory that will be able to float in the clouds for an extended period. Amongst the planned instruments for this payload are not only mass spectrometers but also a microscope that will search cloud droplets for evidence of biological cells.

Perhaps the most ambitious mission planned by the VLF team is a sample return mission that will use a balloon to collect samples of cloud droplets and gas, and return these to Earth for detailed laboratory study. If there is in fact life in the clouds of Venus, a mission like this will be necessary to answer fundamental questions about its origin and how it operates. It is perhaps the dream mission in the search for evidence of life on Venus.

\section{Conclusions}
The discovery of phosphine in the atmosphere of Venus has caused some controversy and has renewed discussions about the possibility of life in the planet's clouds. The observational evidence for phosphine has been challenged and examined in detail. The JCMT and ALMA results have so far survived these challenges, and there has been independent {\em in situ} confirmation of the presence of phosphine from the Pioneer Venus LNMS instrument. Observations from other telescopes in search of phosphine have produced rather more mixed results, with several upper limits and one possible detection. However, the apparent disagreement between these different sets of observations may soon be understood in the context of day-night variations in the amount of phosphine above the clouds thanks to photolysis by sunlight.

While the presence of phosphine in the atmosphere of Venus is becoming more secure with the arrival of new and improved datasets such as JCMT-Venus, an understanding of its origin still eludes us. It is clear that no conventional chemical process can produce phosphine in the amounts observed, but it is still far from clear whether biological processes are involved, or if there is some as-yet unknown non-biological source.

More data is clearly necessary for us to understand what is really going on in the atmosphere of Venus, and this is being sought by a number of different ground and space-based approaches. Over the next several years our understanding of the origin of phosphine on Venus will certainly improve, and we will hopefully reach a point at which the question of life in the clouds of Venus moves from being something that we can only speculate about, to something about which we have clear and decisive knowledge. Whatever conclusion we finally reach, we will have learnt a lot more about our nearest neighbour planet, and this knowledge will help guide our search for possible biospheres on planets orbiting other stars. Confirmation that there is in fact life in the clouds of Venus would be a truly epoch making discovery, but we are still a very long way from drawing that conclusion.
~\\~\\
{\bf Acknowledgements} It is a pleasure to thank Jane Greaves, Janusz Petkowski, Anita Richards, and Wei Tang for many useful comments. It is also a pleasure to thank all members of the phosphine team for their enthusiasm and expertise in what has already been quite an exciting, and unexpected, adventure, which, for me, started in a bar in Hilo.

\end{document}